\documentclass[12pt]{article}
\usepackage{amscd,amsmath,amssymb,amstext,amsthm,exscale,latexsym}
\usepackage{array,graphicx}
\newcommand {\al}   {\alpha}       \newcommand {\bt}  {\beta}
\newcommand {\g }   {\gamma}       
\newcommand {\dl}   {\delta}

\newcommand {\vf }  {\varphi}

\newcommand {\pl}   {\partial}     
\newcommand   {\const}{{\sf\,const}}     
\newcommand {\MR}  {{\mathbb R}}
\begin{document}
\title     {Passing the Einstein--Rosen bridge}
\author    {M. O. Katanaev
            \thanks{E-mail: katanaev@mi.ras.ru}\\ \\
            \sl Steklov Mathematical Institute,\\
            \sl ul. Gubkina, 8, Moscow, 119991, Russia}
\maketitle
\begin{abstract}
We relax the requirement of geodesic completeness of a space-time. Instead, we
require test particles trajectories to be smooth only in the
physical sector. Test particles trajectories for Einstein--Rosen bridge are
proved to be smooth in the physical sector, and particles can freely penetrate
the bridge in both directions.
\end{abstract}
The metric of Einstein--Rosen bridge is \cite{EinRos35}
\begin{equation}                                                  \label{qhhyfd}
  ds^2=g_{\al\bt}dx^\al dx^\bt=\frac{u^2}{u^2+4M}dt^2-(u^2+4M)du^2
  -\frac14(u^2+4M)(d\theta^2+\sin^2\theta d\vf^2),
\end{equation}
where $\al=0,1,2,3$; $x^0=t$, $x^1=u$ are time and space coordinates,
$x^2=\theta$, $x^3=\vf$ are angles of the spherical coordinate system, and $M$
is the Schwarzschild mass. This metric is
smooth on the $t,u$ plane, but degenerate on the line $u=0$. Metric
(\ref{qhhyfd}) glues smoothly two copies of the external Schwarzschild solution
along the line $u=0$ which corresponds to the horizon. One external
Schwarzschild solution is isometric to the half plane $u>0$ multiplied by a
sphere, and the other to the half plane $u<0$.

The well known problem for the Einstein--Rosen bridge is its geodesic
incompleteness at $u=0$. At present, the common point of view is that metric
(\ref{qhhyfd}) satisfies Einstein equations with nontrivial exotic matter
energy--momentum tensor concentrated at $u=0$. This conclusion depends on
how we interpret the singularity at $u=0$.

In the present paper, we propose another point of view. A test particle of mass
$m$ moving in the space-time with metric (\ref{qhhyfd}) is described by the
action
\begin{equation}                                                  \label{epopac}
  S=\int_\g\!d\tau L(q,\dot q)
  :=-m\int_\g\!d\tau\sqrt{g_{\al\bt}\dot q^\al\dot q^\bt},
\end{equation}
where $q^\al(\tau)$, $\tau\in\MR$, is the world line of a test particle. The
action (\ref{epopac}) is reparameterization invariant and therefore is a gauge
model. A test particle has three physical propagating degrees of freedom
corresponding to space coordinates $q^\mu$, $\mu=1,2,3$ (Greek indices from the
middle of the alphabet are assumed to take only space values). The time
coordinate $q^0$ is a gauge degree of freedom, and is not important from the
physical standpoint.

The natural and important requirement to any solution of Einstein equations is
its geodesic completeness (maximal extension along geodesics). In this way, we
obtain the Kruskal--Szekeres extension \cite{Kruska60,Szeker60} of the
Schwarzschild solution which describes black and white holes. Here we relax the
requirement of geodesic completeness. Instead, we require the trajectories of
test particles to be smooth only in the physical sector. We prove that
trajectories of test particles for metric (\ref{qhhyfd}) are smooth and complete
in the physical sector. All singularities are moved to unphysical sector and
can be considered as artifacts of the gauge condition. Thus test particles can
freely penetrate the Einstein--Rosen bridge in both directions.

Action (\ref{epopac}) yields equations of motion (geodesic equations)
\begin{equation}                                                  \label{qgtsju}
  \ddot q^\al=-\Gamma_{\bt\g}{}^\al\dot q^\bt\dot q^\g,
\end{equation}
where $\Gamma_{\bt\g}{}^\al$ are Christoffel's symbols, if we choose the
canonical parameter $\tau$ along the worldline.

\begin{figure}[h,b,t]
\hfill\includegraphics[width=.4\textwidth]{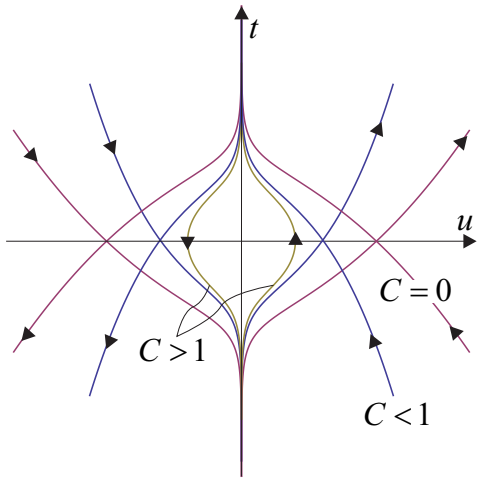}
\hfill {}
\centering\caption{Qualitative behavior of timelike geodesics, $C>0$. Null
geodesics correspond to $C=0$. Every geodesic can be moved along $t$ axis.}
\label{feinrosex}
\end{figure}

We consider only radial geodesics for simplicity. Then geodesic equations
(\ref{qgtsju}) for the Einstein--Rosen bridge reduce to two equations
\begin{align*}
  \ddot t&=-\frac{8M}{u(u^2+4M)}\dot t\dot u,
\\
  \ddot u&=-\frac{4Mu}{(u^2+4M)^3}\dot t^2-\frac u{u^2+4M}\dot u^2.
\end{align*}
They have two integrals of motion
\begin{align*}
  C_0&=g_{\al\bt}\dot x^\al\dot x^\bt=\frac{u^2}{u^2+4M}\dot t^2
  -(u^2+4M)\dot u^2,
\\
  C_1&=K_\al\dot x^\al=\frac{u^2}{u^2+4M}\dot t.
\end{align*}
The last integral is due to the existence of the Killing vector field $K=\pl_0$.
Now we can easily obtain the equation defining the form of radial timelike
geodesics
\begin{equation}                                                  \label{qgkoud}
  \frac{dt}{du}=\pm\frac{u^2+4M}{u\sqrt{1-C\frac{u^2}{u^2+4M}}},
\end{equation}
where $C:=C_0/C_1^2$. The qualitative behavior of timelike geodesics is shown
in Fig.\ref{feinrosex}. They are incomplete at $u=0$.

Strictly speaking, the space-time with metric (\ref{qhhyfd}) is geodesically
incomplete at $u=0$. The requirement that wordlines of test particles be
smooth and defined for all $\tau\in\MR$ in the physical sector yields a natural
way to identify geodesics on different sides of the line $u=0$, thus making the
space-time of Einstein--Rosen bridge geodesically complete. To separate
unphysical degree of freedom we use the well known Hamiltonian formulation
(see, i.e.\ \cite{GitTyu90}). We denote canonically conjugate variable as
$q^\al,p_\al$. The Hamiltonian action for a test particle can be written in the
form
\begin{equation}                                                  \label{epohas}
  S=\int\!dt(p_\al\dot q^\al-H)=\int\!dt\left(p_\mu\dot q^\mu
  -|\dot q^0N|\sqrt{\hat p^2+m^2}+\dot q^0N^\mu p_\mu\right),
\end{equation}
where $N$ and $N^\mu$ are the lapse and shift functions in the ADM
parameterization of the metric
$ds^2=N^2dt^2+g_{\mu\nu}(dx^\mu+N^\mu t)(dx^\nu+N^\nu t)$. We also introduced
notation $\hat p^2:=-\hat g^{\mu\nu}p_\mu p_\nu$ for positive definite square
of space momenta, where $\hat g^{\mu\nu}$ is the three dimensional matrix which
is inverse to $g_{\mu\nu}$. We keep in mind two possible signs of the lapse
function $N>0$ and $N<0$. The zeroes component of momentum is defined by the
first class constraint
\begin{equation}                                                  \label{efipra}
  G:=\sqrt{\hat p^2+m^2}-\left|\frac{p_0-N^\mu p_\mu}N\right|=0.
\end{equation}
The component $q^0$ is a free function to be fixed by the gauge condition.

Consider the case $\dot q^0N>0$. Then Hamiltonian equations of motion for
physical degrees of freedom are
\begin{equation}                                                  \label{qnhfgu}
\begin{split}
  \dot q^\mu&=~~\dot q_0\frac{\pl H_{\rm eff}}{\pl p_\mu},
\\
  \dot p_\mu&=-\dot q^0\frac{\pl H_{\rm eff}}{\pl q^\mu}.
\end{split}
\end{equation}
where we introduced the effective Hamiltonian
\begin{equation*}
  H_{\rm eff}:=N\sqrt{\hat p^2+m^2}-N^\mu p_\mu
\end{equation*}
for physical degrees of freedom.
One can easily check that the energy $E:=H_{\rm eff}$ is conserved for arbitrary
function $q^0(\tau)$. This means that the function $q^0$ describes the freedom
in parameterization of the world line and is to be fixed by a gauge condition.

The lapse function is defined by the metric up to a sign. We choose it to be
\begin{equation}                                                  \label{qjudgr}
  N=\frac u{\sqrt{u^2+4M}}.
\end{equation}
It is positive on the right half plane $u>0$ and negative on the left $u<0$.
This is important, because if we choose positive lapse function $|N|$ everywhere
then it will produce the nontrivial energy-momentum tensor on the right hand
side of Einstein's equations proportional to $\dl(u)$ due to discontinuity in
the first derivative \cite{GuKaNiPa09}, and the Einstein--Rosen solution
(\ref{qhhyfd}) is no longer a solution to the vacuum Einstein's equations. On
the other hand, we proved that the Schwarzschild solution in isotropic
coordinates describes the gravitational field around point particle
\cite{Katana13}. This solution obtained within the Hamiltonian formulation is
globally isometric to Einstein--Rosen bridge, and the lapse function does change
its sign when crossing the critical sphere corresponding to $u=0$.

Now we fix the gauge and analyze equations of motion (\ref{qnhfgu}) for metric
(\ref{qhhyfd}). To simplify notation, we put $q^1:=q$ and $p_1:=p$. Then
equations (\ref{qnhfgu}) take the form
\begin{align}                                                     \label{qglhoy}
  \dot q&=\frac{\dot q^0Np}{(q^2+4M)\sqrt{\hat p^2+m^2}},
\\
  \dot p&=-\dot q^0\sqrt{\hat p^2+m^2}\pl_qN+\frac{\dot q^0Nqp^2}
  {(q^2+4M)^2\sqrt{\hat p^2+m^2}},
\end{align}
where the lapse function is given by Eq.(\ref{qjudgr}) with $u\to q$, and
\begin{equation*}
  \hat p^2=\frac{p^2}{q^2+4M}.
\end{equation*}
The obtained equations can be solved because of the energy conservation:
\begin{equation*}
  E=N\sqrt{\hat p^2+m^2}=\const.
\end{equation*}
From here we can find the momentum
\begin{equation*}
  p=\pm\frac1N\sqrt{(q^2+4M)(E^2-m^2N^2)}.
\end{equation*}
Substituting this solution into Eq.(\ref{qglhoy}) yields
\begin{equation}                                                  \label{qkostr}
  \dot q=\pm\frac{q^0N}{E(q^2+4M)}\sqrt{(q^2+4M)(E^2-m^2N^2)}.
\end{equation}
It defines the form of geodesics $dt/du=\dot q^0/\dot q$ as given by
Eq.(\ref{qgkoud}) where $C=m^2/E^2$. We see that Lagrangian and Hamiltonian
description result in the same geodesic lines as it should be.

\begin{figure}[h,b,t]
\hfill\includegraphics[width=.4\textwidth]{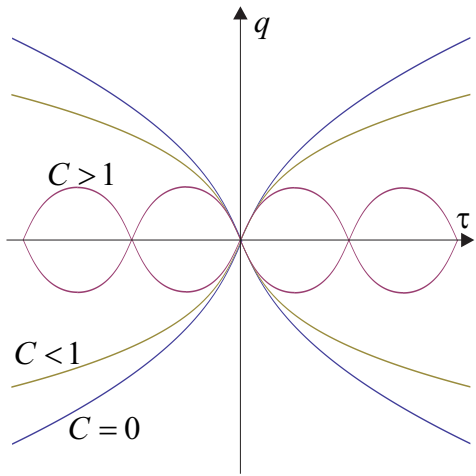}
\hfill {}
\centering\caption{Qualitative behavior of physical trajectories. For $C>1$,
trajectories are oscillating.}
\label{feinrophys}
\end{figure}

Now we fix the gauge $\dot q^0N=1$. Then Eq.(\ref{qkostr}) reduces to
\begin{equation}                                                  \label{qjufsr}
  \dot q=\pm\frac1{\sqrt{q^2+4M}}\sqrt{1-C\frac{q^2}{q^2+4M}}.
\end{equation}
This is the final equation. Its solutions are smooth and defined for all
$\tau\in\MR$. Qualitative behavior of physical trajectories $q(\tau)$ are shown
in Fig.\ref{feinrophys}. Trajectories are oscillating for $C>1$.

At the most interesting point $u=q=0$ Eq.(\ref{qjufsr}) takes the form
$\dot q=\pm1/\sqrt{4M}$ and is nonsingular. We see that a test
particle freely penetrate the line $u=0$ if we exclude unphysical degree of
freedom. From physical point of view this is all that matters, because we may
not bother about the gauge degree of freedom.

In Fig.\ref{feinrosex}, arrows show the direction of increasing $\tau$ which
coincides with increasing of proper time. On the left hand side $\dot q^0<0$,
and the observer sees antiparticle (a particle of the same mass but opposite
charge \cite{GitTyu90}).

The singularity in the Einstein--Rosen metric is avoided because it is moved to
the unphysical sector due to the gauge condition which can be written in the
form
\begin{equation*}
  q^0=\int^\tau d\tau'\frac{q(\tau')^2+4M}{q(\tau')}.
\end{equation*}
It is divergent at $q\to0$ for solutions of Eq.(\ref{qjufsr}).

Physical interpretation of the obtained solution is strange. External observer
at a given moment of time $q^0$ sees simultaneously two particles: a particle on
the right and an antiparticle on the left. For $C>1$, they move simultaneously
either from $u=0$ or towards $u=0$. For $C>1$, the observer sees the birth
of particle and antiparticle at infinite past $q^0=-\infty$ and annihilation of
the pair at infinite future $q^0=\infty$.

Recently, we proved that gravitational field of a point particle is described
by the Schwarzschild solution in isotropic coordinates \cite{Katana13}. This
solution is globally isometric to Einstein--Rosen bridge. The results of the
present paper are also applicable to this case: test particles can freely
penetrate the critical sphere corresponding to $u=0$. In this sense, the
solution is complete and does no require further continuation.

\bibliographystyle{unsrt}
\end{document}